\documentclass[submitting]{nst}

\usepackage{subfigure,dcolumn}
\usepackage[T2A,T1]{fontenc}
\usepackage[russian,english]{babel}

\usepackage{listings}
\usepackage{color}
\begin{document}
% \nolinenumbers

\title{Performance Analysis of Digital Flux-locked Loop Circuit with Different SQUID $V$-$\phi$ Transfer Curves for TES Readout System}
\thanks{This work is supported by the National Key Research
and Development Program of China 
(Grant
No.2021YFC2203400,
No.2022YFC2204900, 
No.2022YFC2204904, 
No.2020YFC2201704),
National Natural Science Foundation of China No.12173040, Youth Innovation Promotion Association CAS 2021011 and Scientific Instrument Developing Project of the Chinese Academy of Sciences,Grant No.YJKYYQ20190065.}

\author{Nan Li}
\affiliation{Key Laboratory of Particle Astrophysics, Institute of High
Energy Physics, CAS, 19B Yuquan Road, Shijingshan District,
100049, Beijing, China}
\author{Le-peng Li}
\affiliation{Key Laboratory of Particle Astrophysics, Institute of High
Energy Physics, CAS, 19B Yuquan Road, Shijingshan District,
100049, Beijing, China}
\author{Meng-jie Song}
\affiliation{Key Laboratory of Particle Astrophysics, Institute of High
Energy Physics, CAS, 19B Yuquan Road, Shijingshan District,
100049, Beijing, China}
\author{Hao-yu Li}
\affiliation{Key Laboratory of Particle Astrophysics, Institute of High
Energy Physics, CAS, 19B Yuquan Road, Shijingshan District,
100049, Beijing, China}
\author{Xiang-xiang Ren}
\affiliation{Key Laboratory of Particle Physics and Particle Irradiation (MOE), Institute of Frontier and Interdisciplinary Science, Shandong University, Qingdao, Shandong, 266237, China}
% \author{Shi-bo Shu}
% \affiliation{Key Laboratory of Particle Astrophysics, Institute of High
% Energy Physics, CAS, 19B Yuquan Road, Shijingshan District,
% 100049, Beijing, China}
% \author{Ya-qiong Li}
% \affiliation{Key Laboratory of Particle Astrophysics, Institute of High
% Energy Physics, CAS, 19B Yuquan Road, Shijingshan District,
% 100049, Beijing, China}
% \author{Yong-jie Zhang}
% \affiliation{Key Laboratory of Particle Astrophysics, Institute of High
% Energy Physics, CAS, 19B Yuquan Road, Shijingshan District,
% 100049, Beijing, China}
% \author{Xu-fang Li}
% \affiliation{Key Laboratory of Particle Astrophysics, Institute of High
% Energy Physics, CAS, 19B Yuquan Road, Shijingshan District,
% 100049, Beijing, China}
\author{Yu-dong Gu}
\affiliation{Key Laboratory of Particle Astrophysics, Institute of High
Energy Physics, CAS, 19B Yuquan Road, Shijingshan District,
100049, Beijing, China}
\author{Cong-zhan Liu}
\affiliation{Key Laboratory of Particle Astrophysics, Institute of High
Energy Physics, CAS, 19B Yuquan Road, Shijingshan District,
100049, Beijing, China}
\author{Hai-feng Li}
\affiliation{Key Laboratory of Particle Physics and Particle Irradiation (MOE), Institute of Frontier and Interdisciplinary Science, Shandong University, Qingdao, Shandong, 266237, China}
\author{He Gao}
\email[Corresponding author, ]{hgao@ihep.ac.cn.}
\affiliation{Key Laboratory of Particle Astrophysics, Institute of High
Energy Physics, CAS, 19B Yuquan Road, Shijingshan District,
100049, Beijing, China}
\author{Zheng-wei Li}
\email[Corresponding author, ]{lizw@ihep.ac.cn}
\affiliation{Key Laboratory of Particle Astrophysics, Institute of High
Energy Physics, CAS, 19B Yuquan Road, Shijingshan District,
100049, Beijing, China}

\begin{abstract}
A superconducting quantum interference device (SQUID), functioning as a nonlinear response device, typically requires the incorporation of a flux-locked loop (FLL) circuit to facilitate linear amplification of the current signal transmitted through a superconducting transition-edge sensor (TES) across a large dynamic range.  
This work presents a reasonable model of the SQUID-FLL readout system, based on a digital proportional-integral-differential (PID) flux negative feedback algorithm. 
This work investigates the effect of $V$-$\phi$ shape on the performance of digital FLL circuits.
% The linear amplification function of the SQUID-FLL system was simulated under different magnetic flux offset conditions. 
% We also analyze the design limits of the bandwidth and slew rate of the system.
Such as the impact factors of bandwidth, design limits of slew rate of the system and the influence of the shapes of SQUID $V$-$\phi$ curve.
Furthermore, the dynamic response of the system to X-ray pulse signals with rise time ranging from $4.4{\sim}281$ $\mathrm{{\mu}s}$ and amplitudes ranging from $6{\sim}8$ $\mathrm{\phi_0}$ was simulated. 
All the simulation results were found to be consistent with the existing mature theories, thereby validating the accuracy of the model. 
The results also provide a reliable modelling reference for the design of digital PID flux negative feedback and multiplexing SQUID readout electronic systems.
\end{abstract}

\keywords{TES, SQUID, Digital flux-locked loop, Readout system.}

\maketitle

\section{Introduction}
As an ultra low-noise detector, a superconducting transition-edge sensor (TES) can achieve an energy resolution ranging from several to several tens of $\mathrm{eV}$~\cite{2409.05643,Liu2024,10068315}. 
The energy resolution of TES is expected to be one to two orders of magnitude better than conventional HPGe semiconductor detectors~\cite{li1998, 1995NEWGR}.
Therefore, TES can be applied in a variety of nuclear detection fields, such as X-ray/$\gamma$-ray detection~\cite{gottardi2021review, ullom2015review, shuo2021development, wang2017generation, uhlig2015high, doriese2017practical, zhang2022tes, fowler2021absolute}, dark matter~\cite{cang2020probing,camus2024cute}, double beta decay~\cite{bratrud2024first}, cosmic microwave background (CMB)~\cite{li2017tibet, cang2020probing}, baryonic matter~\cite{cui2020hubs}, nuclear medicine~\cite{xie2011determination, lin2004dose}, etc.
% In X-ray and $\gamma$-ray detection, TES ,,,
% Even in nuclear medicine, TES detectors can also be used
% to detect $\gamma$-rays from radioactive materials [36, 37].
% It is widely used in detection fields such as $\gamma$ rays~\cite{Keller2024}, X-rays~\cite{2409.05643}, far-IR~\cite{Styers2024}, submillimeter/millimeter waves~\cite{Jaehnig2020}, neutrinos~\cite{singh2023large}, and dark matter detection~\cite{schwemmbauer2023direct}. 
To ensure a high signal-to-noise ratio of TES, a low-noise direct current superconducting quantum interference device (SQUID) is typically used to couple the TES current and then readout~\cite{wu2022development}. 
% A common approach is to use a SQUID to linearly amplify the TES signals.
However, SQUID is a nonlinear device that can only guarantee linear amplification of the signal over a very small dynamic range. 
To enable the nonlinear SQUID to amplify the TES signal linearly over a larger dynamic range, it is necessary to construct a flux negative feedback lock loop (FLL)~\cite{drung2006low}.

A significant benefit of employing a digital system to control a SQUID is the convenient manipulation of digitized signals.
% without the necessity for additional hardware instruments.
For example, the analog integrator and its reset circuit can be easily set up using just a few lines of code~\cite{10.1063/1.1350646, ludwig2001versatile}.
The reset circuit not only automatically resets the position of the $V$-$\phi$ curve prior to the SQUID lock, but also automatically adjusts the position of the FLL circuit working (locking) point.
It has been demonstrated that digital FLL has the capacity to increase the dynamic input range of a SQUID to a substantial degree by using flux-quanta counting and dynamic field
compensation (DFC) methods~\cite{drung2003high}.
The noise spectral density measurement shows that the digital FLL system performs in a similar way to the analog FLL controller in previous work~\cite{LIMKETKAI20031506}.
The digital FLL is now widely used in experiments such as magnetoencephalogram (MEG)~\cite{VRBA2001249}, magnetocardiogram (MCG)~\cite{oyama2006development}, CMB~\cite{CMB-S4:2016ple}, X-ray and $\gamma$-ray detection~\cite{Barret2023,Zhang2022}.
% The focus of these digital FLL experiments is on system noise, with a lack of detailed discussion of Influencing factors for digital FLL system performance.
Digital FLL is used in many time-division multiplexing (TDM) systems for TES readout and have detailed description of the digital feedback logic\cite{reintsema2003prototype}.
% For example, it is necessary to have a large signal dynamic range~\cite{oyama2006development} in MEG and MCG experiments 
It is necessary to have fast response in X-ray detection, which requires the FLL to have a large slew rate\cite{durkin2020predictive}, typically several $\mathrm{{\mu}s}$ at 10 $\mathrm{{\mu}A}$ ~\cite{kilbourne2007uniform}, which is less than 2 $\mathrm{{\phi}_0}$ if the mutual inductance coefficient of SQUID coil is 5 $\mathrm{{\mu}A/{\phi}_0}$. In CMB detection experiments, the design of digital FLL requires large bandwidth in order to improve multiplexing numbers of readout channels~\cite{Doriese2016}.
In order to understand the performance of the digital SQUID-FLL system and to design higher performance readout circuits, we simulate the functional upper limits of SQUID-FLL system.
% \textcolor{blue}{Previous works has investigated the effect of different shapes of SQUID $V$-$\phi$ curves on the performance of digital FLL systems.}
We also analyze the effect of shapes on the performance of digital FLL using three different shapes of SQUID $V$-$\phi$ curve models.  
% Thus the digital FLL circuit is simulated for different SQUID flux response using the circuit model of SQUID-FLL.

The paper is organized as follows: 
% This model provides a reasonable simplification of the transfer function of the nonlinear SQUID for the input current (or magnetic flux) signal, as discussed in Section~\ref{sec:SQUID_model}. 
In Section~\ref{sec:SQUID_FLL_princp}, a sinusoidal-like approximation is made to the SQUID flux response model, and the fundamentals of the digital SQUID-FLL circuit are introduced.
% A comprehensive FLL circuit model was constructed based on a digital proportional integral differential (PID) algorithm, as describe in Section~\ref{sec:FLL_model}. 
In Section~\ref{sec:FLL_sim}, we conduct a comprehensive simulation analysis to validate the reliability of the model.
This primarily entails simulations of the fundamental linear amplification readout function of the input magnetic flux signal, system bandwidth, and slew rate.  
% The simulation results were compared and validated against established and mature theoretical models, which 
Finally, the dynamic response of the system was simulated using pulse signals that were designed to resemble actual TES pulse signals with varying rise time and amplitudes.
This SQUID-FLL readout system model can provide reliable design references for digital flux negative feedback (DFB) readout and multiplexing SQUID readout systems.
Section~\ref{sec:conclusion} concludes the paper.

\section{Principle of digital SQUID-FLL model}
\label{sec:SQUID_FLL_princp}
\subsection{Simplified SQUID transfer model}
\label{sec:SQUID_model}
\begin{figure*}[!htb]
    \centering
    \includegraphics[width=.9\textwidth]{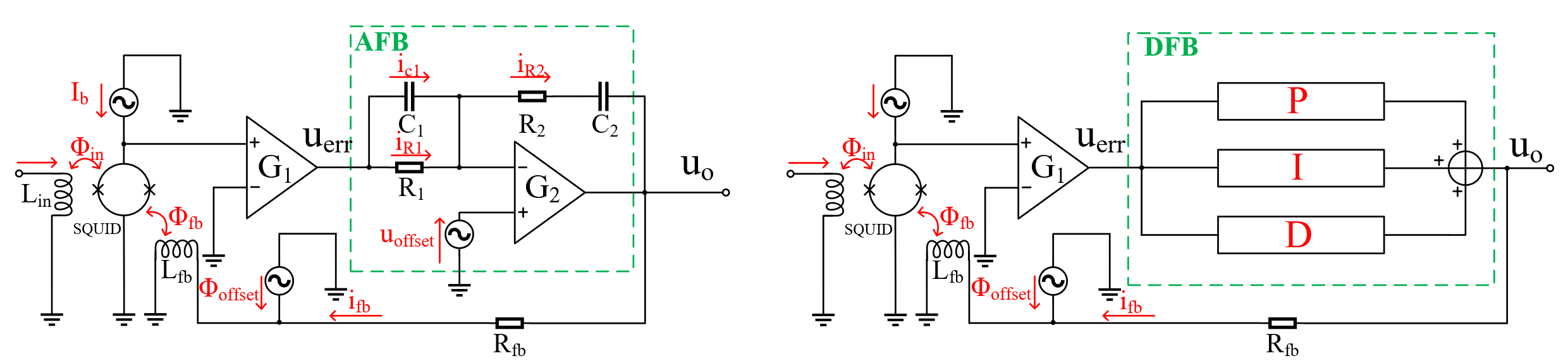}
    % \qquad
    % \includegraphics[width=.45\textwidth]{figures/fig1_DFB_2.png}
    \caption{SQUID-FLL readout circuit model. $\phi_{in}$ is the input flux signal and $\phi_{fb}$ is the feedback flux. $G_1$ is linear gain of room temperature preamplifier. $\phi_{offset}$ is the magnetic flux offset bias applied to the feedback coil here. $u_{offset}$ is the voltage offset of the amplifier. 
    The left figure shows the principle design of analog SQUID-FLL circuit. The right figure is the equivalent digital PID circuit. Analog feedback (AFB) and digital feedback (DFB) circuits are included in the green dashed lines.\label{fig:PID_design}}
\end{figure*}

The circuit model of the SQUID-FLL is illustrated in Figure~\ref{fig:PID_design}.
It is comprised of a single-stage SQUID.
The current signal flowing through the TES is coupled with the input coil $L_{in}$ and converted into a magnetic flux signal $\phi_{in}$.
Although different designs for the SQUIDs exhibit varying responses to input magnetic flux $\phi_{in}$, and all have common characteristics: the response of SQUID to $\phi_{in}$ is periodic and nonlinear, which resembles a quasi-sinusoidal curve~\cite{doi:https://doi.org/10.1002/3527603646.ch4}. 
Following magnetic flux locking, the error between the input and feedback magnetic flux is nearly zero, maintaining the total flux value in the SQUID at the working point. 
Theoretically, we only need to know the linear conversion coefficient of $dV_{SQUID}/d\phi_{in}$ at the lock point~\cite{DRUNG2002134} but do not need to care any shape of the $V$-$\phi$ response curve of SQUID, where $V_{SQUID}$ represents the voltage across SQUID. 
To prove this point, we simulate the FLL circuit for different SQUID flux response cases in Section~\ref{sec:FLL_sim}.
% Subsequently, the SQUID-FLL model was simulated accurately.  
% Therefore, despite the fact that the SQUID response to $\phi_{in}$ is not a perfect sine wave, it is still possible to

We simplify the SQUID $V$-$\phi$ transfer model to a standard sinusoidal function.
Assuming that the mutual inductance coefficient of the SQUID input coil is $M_{in}$ and the mutual inductance coefficient of the feedback coil is $M_{fb}$,
the relationship between the error signal $u_{\text{err}}(t)$ and the input and feedback magnetic fluxes can be expressed as follows:
\begin{widetext}
\begin{equation}
\label{eq:0}
% \left\{
    u_{err}(t) = Asin\left(\frac{2\pi}{\phi_{0}}\left[i_{in}(t)M_{in} - \frac{u_{o}(t)M_{fb}}{R_{fb}}+\phi_{offset}\right]\right)G_{1}
% \right.
\end{equation}
\end{widetext}
where $\phi_{0}$ denotes the quantum magnetic flux. 
$i_{in}(t)$ is the TES signal through the input coil and the input flux is given by $\phi_{in}=i_{in}(t)M_{in}$. 
$\phi_{offset}$ is the magnetic flux bias applied to the input or feedback coils. 
$u_{o}(t)$ denotes the feedback voltage signal. 
$R_{fb}$ denotes the feedback resistor and the feedback magnetic flux is given by $\phi_{fb}=\frac{u_{o}(t)M_{fb}}{R_{fb}}$.
$G_{1}$ denotes the linear gain of the room low-noise preamplifier.
$A$ denotes the amplitude of the voltage across the SQUID.
An offset voltage $u_{offset}$ can also be introduced into the circuit to bias the preamplifier when tuning the lock point, as shown in Figure~\ref{fig:PID_design}.
% $u_{offset}$ the voltage offset of the amplifier.

\subsection{Principle of digital FLL}
\label{sec:FLL_model}
The fundamental structure of the FLL is the construction of a closed-loop negative flux feedback circuit based on operational amplifiers~\cite{drung2001improved}, as shown in the left part of Figure~\ref{fig:PID_design}.
By analyzing the current relationships contained within the green dashed line in Figure~\ref{fig:PID_design}, we can derive the circuit equation relating the output digtial signal $u_{o}(n)$ to the input error signal $u_{err}(n)$:
% \begin{equation}
% \label{eq:1}
% % \left\{
%     \begin{cases}    
%            \displaystyle \frac{u_{err}(t)}{R_1} + C_1\frac{du_{err}(t)}{dt} &= \displaystyle-C_2\frac{du_{C_{2}}(t)}{dt} \\
%            \displaystyle u_{R_2}(t) + u_{C_2}(t) &=\displaystyle u_o(t) \\
%            \displaystyle \frac{u_{R_2}(t)}{R_2} &=\displaystyle C_2\frac{du_{C_2}(t)}{dt}       
%     \end{cases}  
% % \right.
% \end{equation}
\begin{widetext}
\begin{equation}
\label{eq:3}
% \left\{
    u_{o}(n) = -\left\{\left(\frac{R_2}{R_1} + \frac{C_1}{C_2}\right)u_{err}(n) + \frac{\Delta{t}}{R_{1}C_{2}}\sum_{i=0}^{n}{u_{err}(i)} + \frac{R_{2}C_{1}}{\Delta{t}}[u_{err}(n) - u_{err}(n-1) ]\right\}
% \right.
\end{equation}
\end{widetext}
where $R_1$ and $C_1$ are the input resistance and capacitance, respectively. $R_2$ and $C_2$ are the feedback resistance and capacitance, respectively. 
% $u_{R}(t)$ and $u_{C}(t)$ are the corresponding voltages. 
$P{\equiv}\left(\frac{R_2}{R_1} + \frac{C_1}{C_2}\right)$ is a proportional parameter. 
$I{\equiv}\frac{\Delta{t}}{R_{1}C_{2}}$ is the integral parameter.
$D{\equiv}\frac{R_{2}C_{1}}{\Delta{t}}$ is the differential parameter.
$\Delta{t}$ represents the digital sampling period.
$u_{err}(i)$ denotes the $i$-th sampled error signal after amplification by $G_1$.
$u_{err}(n)$ is the average value of all sampled $u_{err}(i)$ in one frame as shown in Figure~\ref{fig:Digital_time_logic}.
$u_{o}(n)$ is the $n$-th output (feedback) signal.
% So we can obtain:
% \begin{equation}
% \label{eq:2}
% % \left\{
%     u_{o}(t) = \left(\frac{R_2}{R_1} + \frac{C_1}{C_2}\right)u_{err}(t) + \frac{1}{R_{1}C_{2}}\int{u_{err}(t)dt} + R_{2}C_{1}\frac{du_{err}(t)}{dt}
% % \right.
% \end{equation}
% Therefore a digital equivalent circuit can be obtained, as illustrated by the digital feedback (DFB) indicated by the green dashed lines in the right part of Figure~\ref{fig:PID_design}. 
% This leads to the following digital PID model:
% \begin{equation}
% \label{eq:3}
% % \left\{
%     u_{o}(n) = \left(\frac{R_2}{R_1} + \frac{C_1}{C_2}\right)u_{err}(n) + \frac{\Delta{t}}{R_{1}C_{2}}\sum_{i=0}^{n}{u_{err}(i)} + \frac{R_{2}C_{1}}{\Delta{t}}[u_{err}(n) - u_{err}(n-1) ]
% % \right.
% \end{equation}
% where $P{\equiv}\left(\frac{R_2}{R_1} + \frac{C_1}{C_2}\right)$ is a proportional parameter. 
% $I{\equiv}\frac{\Delta{t}}{R_{1}C_{2}}$ is the integral parameter.
% $D{\equiv}\frac{R_{2}C_{1}}{\Delta{t}}$ is the differential parameter.
% $\Delta{t}$ represents the digital sampling period.
% $u_{err}(i)$ denotes the $i$-th sampled error signal after amplification by $G_1$.
% $u_{err}(n)$ is the average value of all sampled $u_{err}(i)$ in one frame as shown in Figure~\ref{fig:Digital_time_logic}.
% $u_{o}(n)$ is the $n$-th output (feedback) signal.
Similarly, by digitizing Equation~\ref{eq:0}, we obtain:
\begin{widetext}
\begin{equation}
\label{eq:4}
% \left\{
    u_{err}(i) = Asin\left(\frac{2\pi}{\phi_{0}}\left[i_{in}(i)M_{in} - \frac{u_{o}(n)M_{fb}}{R_{fb}}+\phi_{offset}\right]\right)G_{1}
% \right.
\end{equation}
\end{widetext}

The primary objective of the PID algorithm design is to calibration of the three aforementioned parameters. If these parameters are not adequately calibrated, system oscillations can occur, which may lead to an extended time to lock (track) the signal or even a loss of lock.

% \section{Simulation results of digital SQUID-FLL models}
\section{Digital SQUID-FLL performance analysis}
\label{sec:FLL_sim}

\subsection{Basic characterization of digital SQUID-FLL system}
In our simulation, we emulated the typical conditions employed in laboratory tests, assuming that the SQUID is biased at a position with a substantial voltage swing prior to flux locking. 
The current through TES $i_{{in}}(i)$ serves as the input reference for the SQUID input coil $L_{in}$. 
The default parameters used in the simulation, unless otherwise specified, are listed in Table~\ref{tab:1}.
\begin{table}[htbp]
    \centering
    \caption{Default simulation parameters of SQUID-FLL system.\label{tab:1}}
    % \smallskip
    % \resizebox{0.6\textwidth}{0.1\textwidth}{
    \begin{tabular}{c|c}
        \hline
        $Parameter$ & $Value$ \\
        \hline
        Input signal type & Triangle Wave \\
        Input signal frequency & $23$ {$\mathrm{Hz}$} \\
        Input signal amplitude & $50$ {$\mathrm{{\mu}A}$} \\
        Testing time & $50$ {$\mathrm{ms}$} \\
        Digital sample rate & $150$ {$\mathrm{MSPS}$} \\
        Sample numbers per feedback & $7$ \\
        Mutual inductance of input coil $1/M_{in}$ & {$28$ $\mathrm{{\mu}A/\phi_0}$} \\
        Mutual inductance of feedback coil $1/M_{fb}$ & {$38$ $\mathrm{{\mu}A/\phi_0}$} \\
        Conversion coefficient $dV_{SQUID}/d\phi_{in}$ at lock point & $3$ {$\mathrm{mV/\phi_0}$} \\
        Input resistor $R_1$ & $100$ {$\mathrm{\Omega}$} \\
        Feedback resistor $R_2$ & $50$ {$\mathrm{\Omega}$} \\
        Input capacitor $C_1$ & $0$ {$\mathrm{nF}$} \\
        Feedback capacitor $C_2$ & $1$ {$\mathrm{nF}$} \\
        Feedback resistor $R_{fb}$ & $10$ {$\mathrm{k\Omega}$} \\
        Preamplifier gain $G_1$ & $100$ \\
        Voltage amplitude across SQUID Array & $700$ {$\mathrm{{\mu}V}$} \\
        Preamplifier voltage offset $u_{offset}$ & $0$ {$\mathrm{{\mu}V}$} \\
        Flux offset $\phi_{offset}$ & $0$ {$\mathrm{{\mu}\phi_0}$} \\
        Settle time $t_{set}$ & $0$ {$\mathrm{ns}$} \\
        \hline
    \end{tabular}
    % }
\end{table} 

\begin{figure*}[!htb]
    \centering
    \includegraphics[width=.9\textwidth]{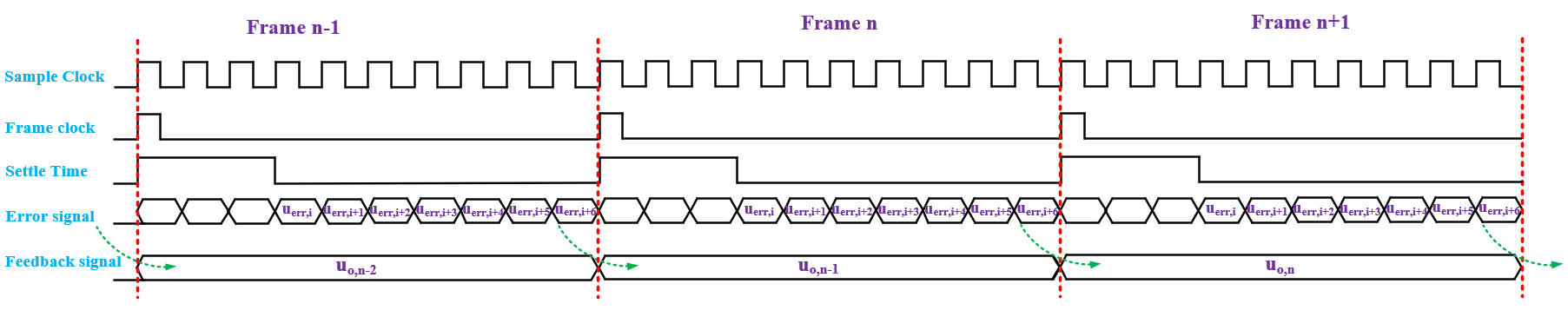}
    \caption{Time logic of digital sample frequency and frame clock (feedback frequency). Sampled error signals $u_{err}(i)$ are used to calculate next frame feedback signal $u_{o}(n)$ based on PID algorithm.\label{fig:Digital_time_logic}}
\end{figure*}

\begin{figure}[htbp]
    \centering
    \includegraphics[width=.45\textwidth]{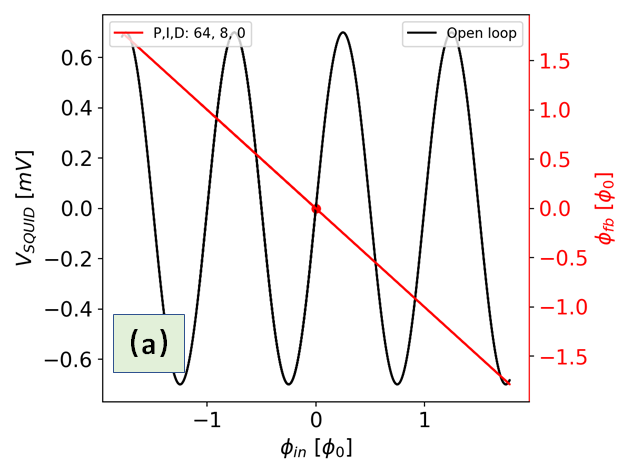}
    \qquad
    \includegraphics[width=.45\textwidth]{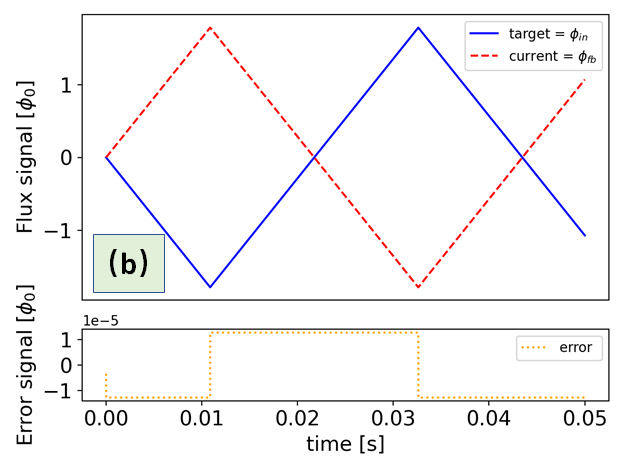}
    \caption{Simulation result of SQUID-FLL model based on PID algorithm with low frequency triangle wave. (a) Black line show the relationship between output flux $\phi_{fb}$ and input flux $\phi_{in}$ without flux lock loop. Red line shows the locked linear gain with default PID parameters at lock point (red dot). All PID parameters on the legend are relative values calculated according to Table~\ref{tab:1} and shifted 7 bits to the left. (b) $\phi_{fb}$ (red dashed line) tracks $\phi_{in}$ (blue solid line) in time domain after flux lock. The orange dotted line below shows the tracked flux error, which is less than 10 $\mathrm{\mu{\phi_0}}$. \label{fig:sim_FLL_lock}}
\end{figure}

% Owing to the low thermal conductance $G$ of TES~\cite{TES_in_CPD}, 
The rise time of TES's signals are typically from several ${\mu}s$ to several tens of $ms$~\cite{THURGATE2021165707,Yagi2023}.
Consequently, the feedback algorithm design does not require consideration of future trend predictions.
Consequently, in the default settings, the input capacitance $C_1$ was set to $0$, retaining only the $P$ and $I$ components. 
% The test time is set to $50$ $ms$ to ensure at least one cycle of the input signal.
Considering the prospective system design, the digital sampling time was established at 150 MSPS, with seven points sampled in each frame.
The frame clock (feedback frequency) was approximately 21.4 MHz, and the effective system Nyquist bandwidth is approximately 10.7 MHz.
In digital feedback, the error signal from the current frame is used to compute the feedback signal for the next frame, which results in a delayed feedback loop. 
This process is analogous to TDM SQUID readout logic, with the exception of row selection~\cite{reintsema2003prototype, battistelli2008functional}. 
In this configuration, the frame frequency is equal to the row selection frequency in the TDM.
The timing logic is illustrated in Figure~\ref{fig:Digital_time_logic}.
Average value of the error signal, $u_{err}(n)$, is calculated from the $u_{err}(i)$ to $u_{err}(i+6)$ values in the $n$-th frame. The feedback signal $u_{o}(n)$ is then calculated for the next frame.
It should be noted that our model does not take into account other system delays, which typically amount to several hundred nanoseconds and may include transmission line delays.  
While these delays may reduce the system bandwidth, they do not impact the model's behaviour.
Prior to the locking of the SQUID, it is essential to adjust the locking work point. 
Typically, in order to ensure linear amplification with no offset, the lock point is selected at the zero point of the input flux ($\phi_{in} = 0$) and the zero point of the output (feedback) flux ($\phi_{fb} = 0$).
The initial error signal is set to 0. The bias signals $\phi_{offset}$, $u_{offset}$ and settle time $t_{set}$ are also set to 0. 
The results of the feedback locking are shown in Figure~\ref{fig:sim_FLL_lock}. The absolute error signal does not exceed 10 $\mathrm{\mu{\phi_0}}$, and the calculated relative error signal is less than 0.1 $\mathrm{\%}$.

\begin{figure}[htbp]
    \centering
    \includegraphics[width=.455\textwidth]{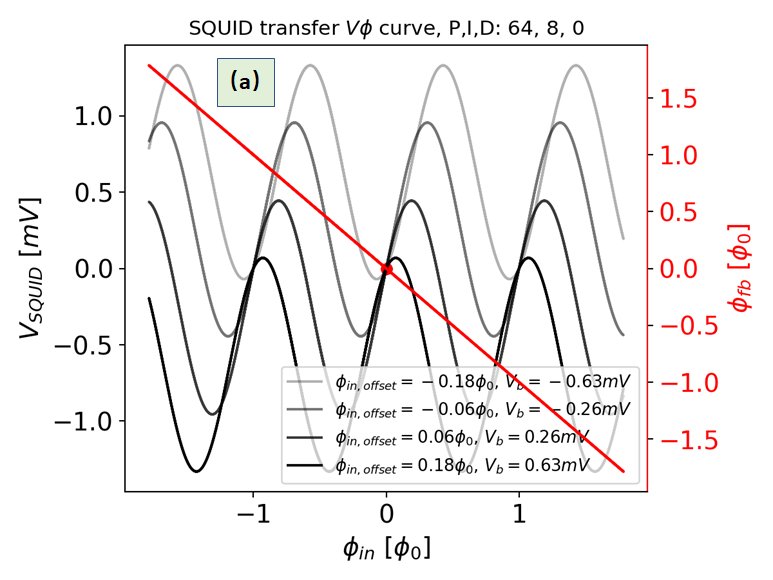}
    % \quad
    \includegraphics[width=.44\textwidth]{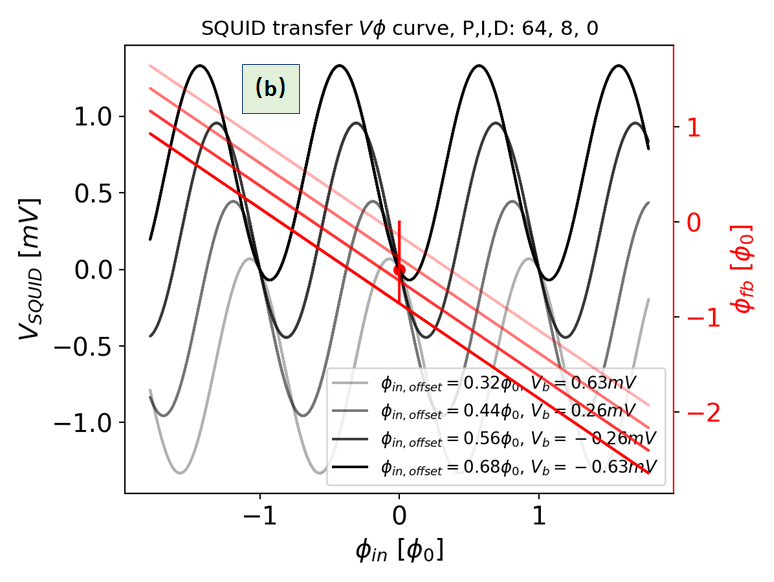}
    % \quad
    \includegraphics[width=.47\textwidth]{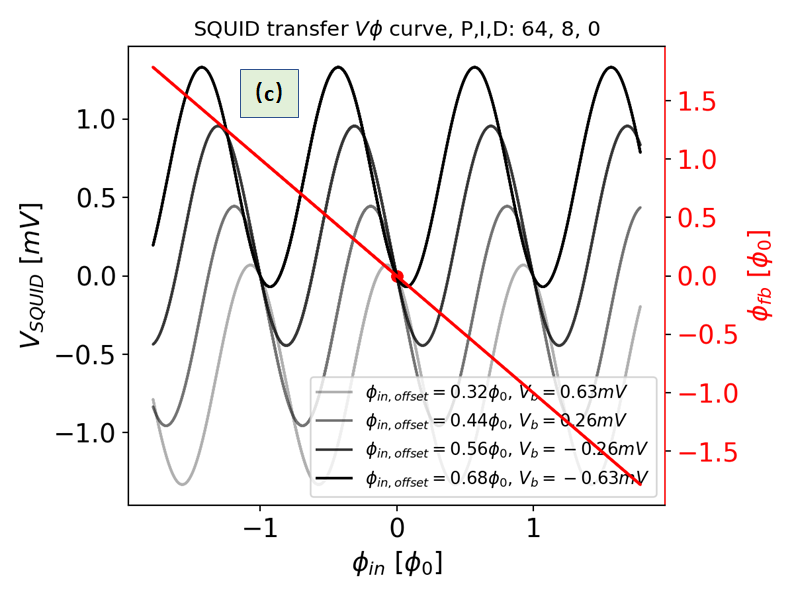}
    \caption{Simulation results of SQUID-FLL model (red lines) under different flux offsets $\phi_{offset}$ and voltage offset $u_{offset}$ (black lines). $u_{offset}$ is added after $G_1$.} (a) Results at different flux offset when room amplification gain is negative and $dV_{SQUID}/d\phi_{in}>0$ at locking point. (b) Room amplification gain is negative but $dV_{SQUID}/d\phi_{in}<0$ at locking point. Locking point offset appears. (c) Room amplification gain is positive and $dV_{SQUID}/d\phi_{in}<0$ at locking point.\label{fig:sim_FLL_lock_diff_offset}
\end{figure}

\begin{figure}[htbp]
    \centering
    \includegraphics[width=.45\textwidth]{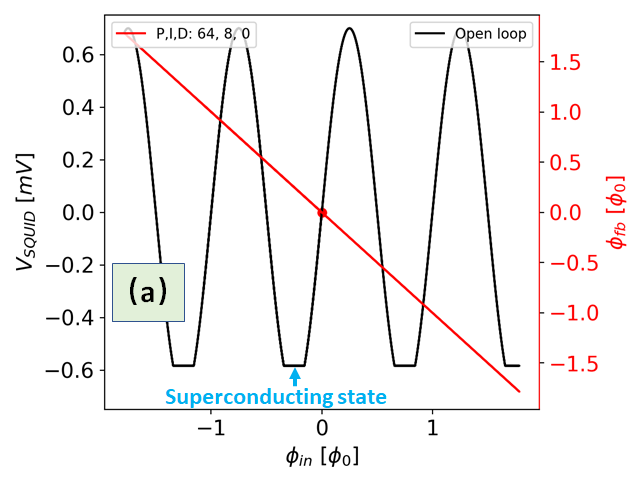}
    % \qquad
    % \includegraphics[width=.45\textwidth]{figures/fig3_sim_FLL_lock_XY_non_sin.png}
    % \qquad
    \includegraphics[width=.45\textwidth]{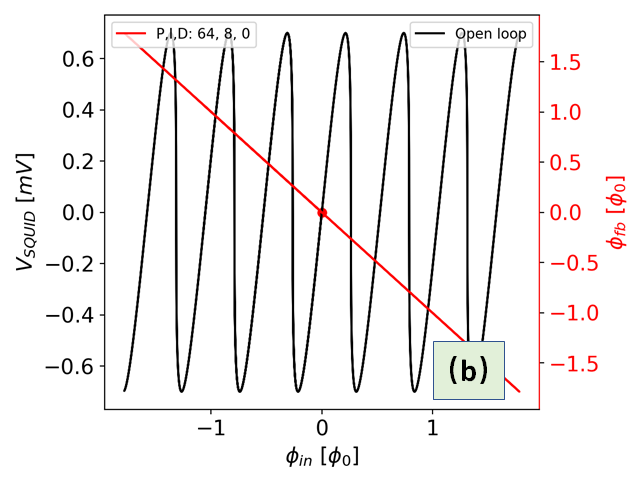}
    \caption{Simulation results of SQUID-FLL (red lines) under varying SQUID $V$-$\phi$ response. (a) SQUID lock status at small bias current. The superconducting state exists at this moment. (b) Lock status at asymmetric response. \label{fig:sim_FLL_lock_diff_vphi}}
\end{figure}

The locking conditions were simulated under different flux and voltage offset conditions. 
In typical circumstances, the response of the input flux $dV_{SQUID}/d\phi_{in}$, near the SQUID locking point, must have the opposite polarity as the room temperature amplifier gain to ensure that the SQUID locking point is maintained. This is shown in Figures~\ref{fig:sim_FLL_lock_diff_offset} (a) and (c).
In the event that the polarity of the room temperature amplifier gain is identical to that of the SQUID response, a locking point offset phenomenon is observed, as illustrated in Figure~\ref{fig:sim_FLL_lock_diff_offset} (b).
In order to verify the independence of the FLL circuit from the shape of the SQUID $V$-$\phi$ response curve, as illustrated in Section~\ref{sec:SQUID_model}, we simulate flux locking for different shapes of the SQUID $V$-$\phi$ curve, as shown in Figure~\ref{fig:sim_FLL_lock_diff_vphi}.

\subsection{Bandwidth and slew rate of digital SQUID-FLL system}
\begin{figure}[htbp]
    \centering
    \includegraphics[width=.45\textwidth]{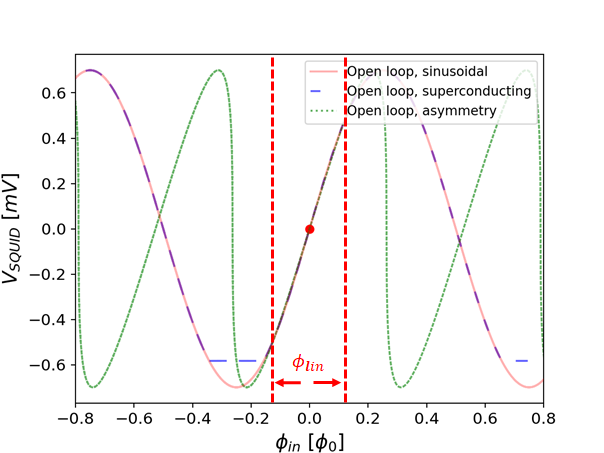}
    \includegraphics[width=.46\textwidth]{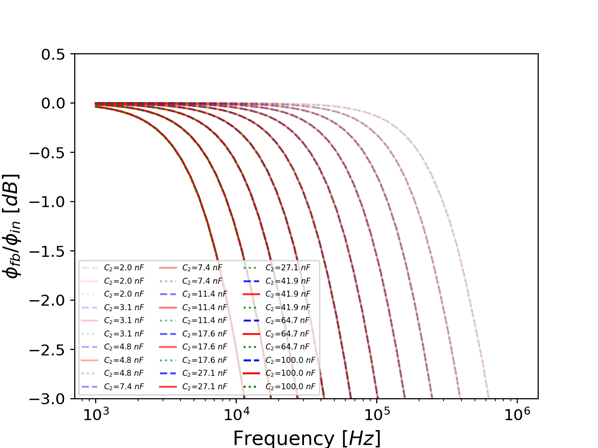}
    \caption{SQUID-FLL system bandwidth simulation. The top figure shows the same open-loop linear gain input range $\phi_{lin}$ for three different shapes of $V$-$\phi$ response. The red solid line is the sinusoidal-like model. The blue dashed line is the small bias model with superconducting state. The green dotted line is the asymmetrical model. The bottom figure shows the bandwidth with different feedback capacitor $C_2$ for three varying shapes of $V$-$\phi$ curves, which prove that $V$-$\phi$ responses of different shapes have equivalent bandwidth when they have same $V_{\phi}$ conversion coefficient at the locking point.\label{fig:sim_FLL_lock_bandwidth}}
\end{figure}

\begin{figure}[htbp]
    \centering
    \includegraphics[width=.45\textwidth]{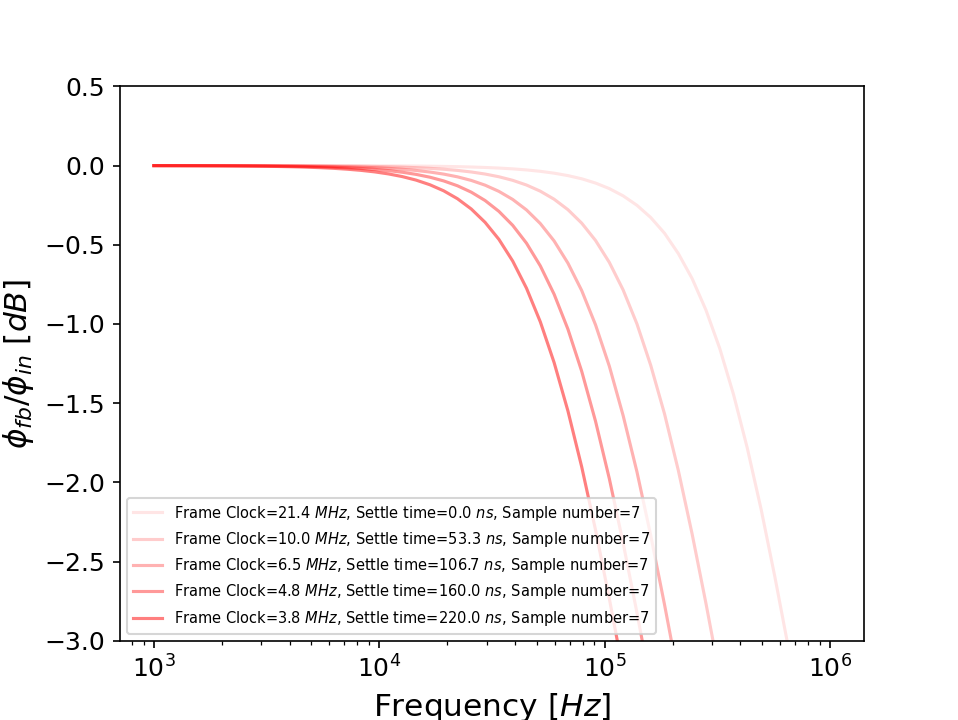}
    % \qquad
    \includegraphics[width=.45\textwidth]{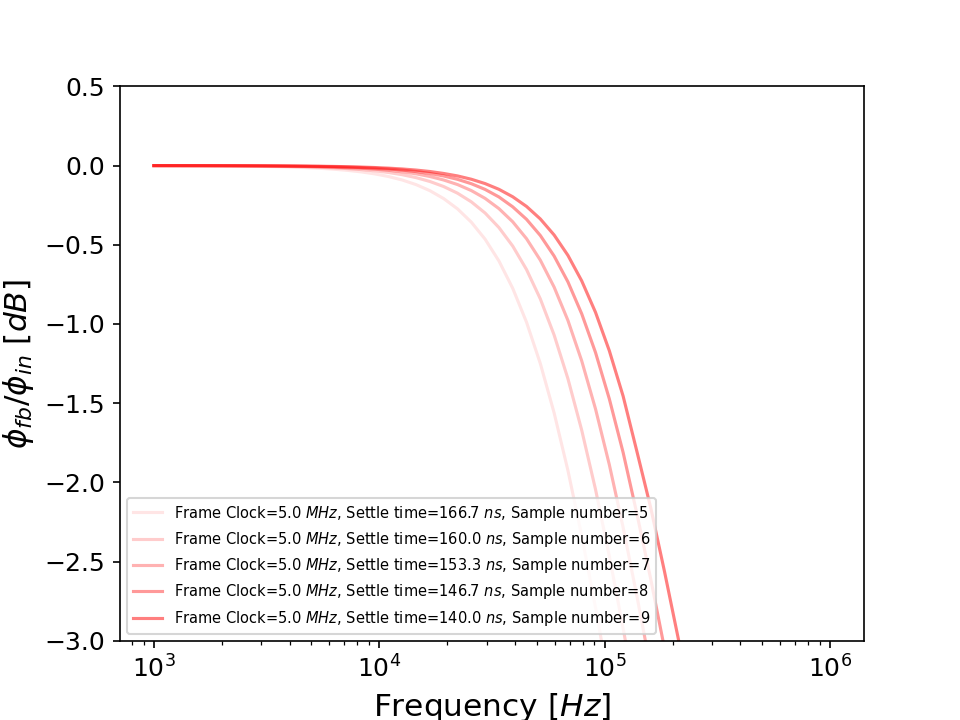}
    \caption{Simulation results of SQUID-FLL under varying settle time. 
    The top figure shows the bandwidth increases with the frame rate when $N_{sp}$ is constant.
    The bottom figure shows the bandwidth increases with $N_{sp}$ when the frame rate is constant.
     \label{fig:sim_FLL_lock_diff_frame_clk}}
\end{figure}

When designing a SQUID-FLL circuit, it is essential to consider the rational design of the system bandwidth, particularly in multiplexing experiments. 
An FLL with a small bandwidth may result in distortion of the TES signal, whereas a larger bandwidth could introduce additional high-frequency noise, which would in turn affect the energy resolution of the TES.
The effect of different feedback capacitors $C_2$ (integral component) and different shapes of $V$-$\phi$ transfer curves on the bandwidth was simulated, and the results are shown in Figure~\ref{fig:sim_FLL_lock_bandwidth}.
In Figure~\ref{fig:sim_FLL_lock_bandwidth}, we consider the ideal case, without considering digital circuit delays ($t_{set}$ $=$ $0$), as in analog FLL.
All SQUID $V$-$\phi$ curves have same conversion coefficient $dV_{SQUID}/d\phi_{in}$ at the locking point and linear dynamic input range.
The PID feedback is a time-domain simulation. By comparing it with the frequency-domain transfer function of the system, the correctness of the SQUID-FLL models can be verified.
Based on frequency-domain transfer function~\cite{doi:https://doi.org/10.1002/3527603646.ch4}, the theoretical 3 dB system bandwidth can be obtained:
\begin{equation}
\label{eq:5}
% \left\{
    % \begin{aligned}
        f_{c}{\approx}\frac{V_{\phi}G_{1}M_{fb}}{2{\pi}R_{1}C_{2}R_{fb}} 
    % \end{aligned}
% \right.
\end{equation}
where $V{\phi}=dV_{SQUID}/d\phi_{in}$ denotes the conversion coefficient at the locking point. For example, when $C_2 = 2$ $\mathrm{nF}$, the calculated value for the critical frequency $f_{c}$ based on Equation~\ref{eq:5} is $628$ $\mathrm{kHz}$, which is in accordance with the results of different $V$-$\phi$ transfer models displayed in Figure~\ref{fig:sim_FLL_lock_bandwidth}.
From Equation~\ref{eq:5}, it can be observed that a multitude of factors influence bandwidth, all of which are linearly correlated. 
In room temperature circuit design, the bandwidth can be determined by adjusting the preamp gain $G_1$, input resistance $R_1$, feedback capacitance $C_2$, and feedback resistance $R_{fb}$.
It is crucial to acknowledge that $G_1$ represents an ideal linear gain in the simulation. $R_2$ affects the DC gain of FLL, which is the discharge resistor for the $C_2$, but does not determine the position of the $f_c$. In an actual design, the limitations of the operational amplifier's gain bandwidth product (GBP) must be considered. Therefore, this gain should not be excessively large for a system with a large bandwidth. Furthermore, in a practical design, system delays also reduce the bandwidth, necessitating the minimization of cable lengths.
If the system delay is considered ($t_{set}$ $\neq$ $0$), the system bandwidth should satisfy Equation~\ref{eq:5.1} and the simulation results are shown in Figure~\ref{fig:sim_FLL_lock_diff_frame_clk}. $t_{sp}$ is the ADC sampling period and $N_{sp}$ is the number of samples. $t_{fm}$ is the frame period.
\begin{equation}
\label{eq:5.1}
% \left\{
    % \begin{aligned}
        f_{c}{\approx}\frac{V_{\phi}G_{1}M_{fb}N_{sp}t_{sp}}{2{\pi}R_{1}C_{2}R_{fb}t_{fm}} 
    % \end{aligned}
% \right.
\end{equation}

\begin{figure}[htbp]
    \centering
    \includegraphics[width=.45\textwidth]{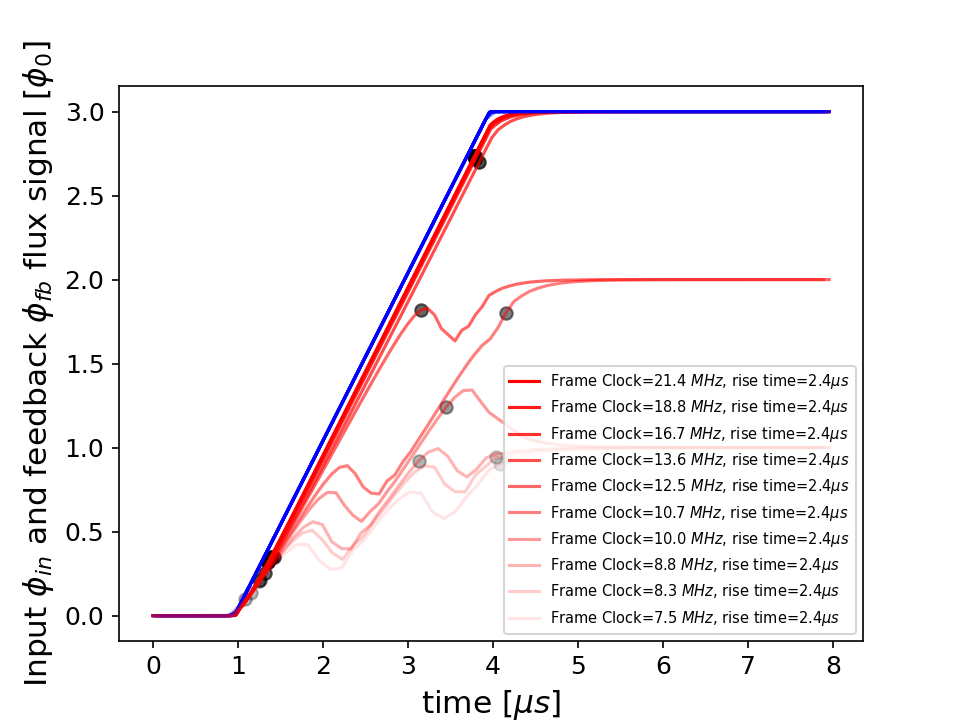}
    \caption{Tracking 2.4 $\mu{s}$, 3 $\phi_0$ signals at different frame rates. The signal is distorted when the frame rate is reduced to 12.5MHz.\label{fig:sim_FLL_lock_frame_clk}}
\end{figure}

\begin{figure}[htbp]
    \centering
    \includegraphics[width=.45\textwidth]{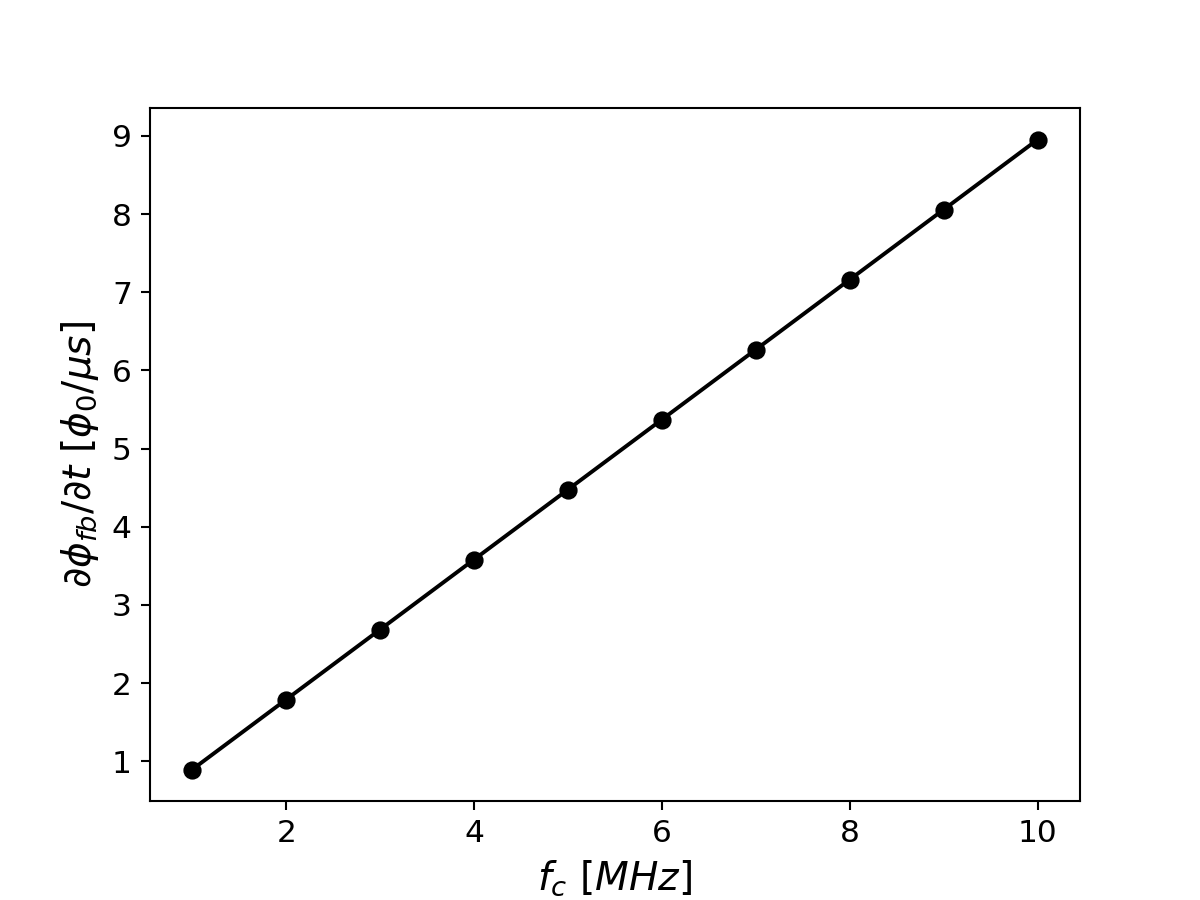}
    \caption{The ideal relationship between SQUID-FLL system slew rate $\partial{\phi_{fb}}/\partial{t}$ and cutoff frequency $f_c$. \label{fig:sim_FLL_lock_slewrate_fc}}
\end{figure}

Further investigation was conducted on the system slew rate $\partial{\phi_{fb}}/\partial{t}$. 
When the bandwidth is sufficiently large but the slew rate is too small, it can also lead to signal distortion.
The system slew rate is determined jointly by the bandwidth, the input range $\phi_{lin}$ of the SQUID's open-loop linear gain, and the design of the FLL integral parameter, and is independent of frequency~\cite{DRUNG2002134}.
This relationship can be expressed as follows:
\begin{equation}
\label{eq:6}
% \left\{
    % \begin{aligned}
        \frac{\partial{\phi_{fb}}}{\partial{t}}=\frac{\phi_{lin}N_{sp}t_{sp}}{2R_{1}C_{2}t_{fm}} 
    % \end{aligned}
% \right.
\end{equation}
In our SQUID transmission model, $\phi_{lin}{\approx}0.3$ $\mathrm{\phi_0}$, as shown on the top of Figure~\ref{fig:sim_FLL_lock_bandwidth}.
\begin{figure}[htbp]
    \centering
    \includegraphics[width=.45\textwidth]{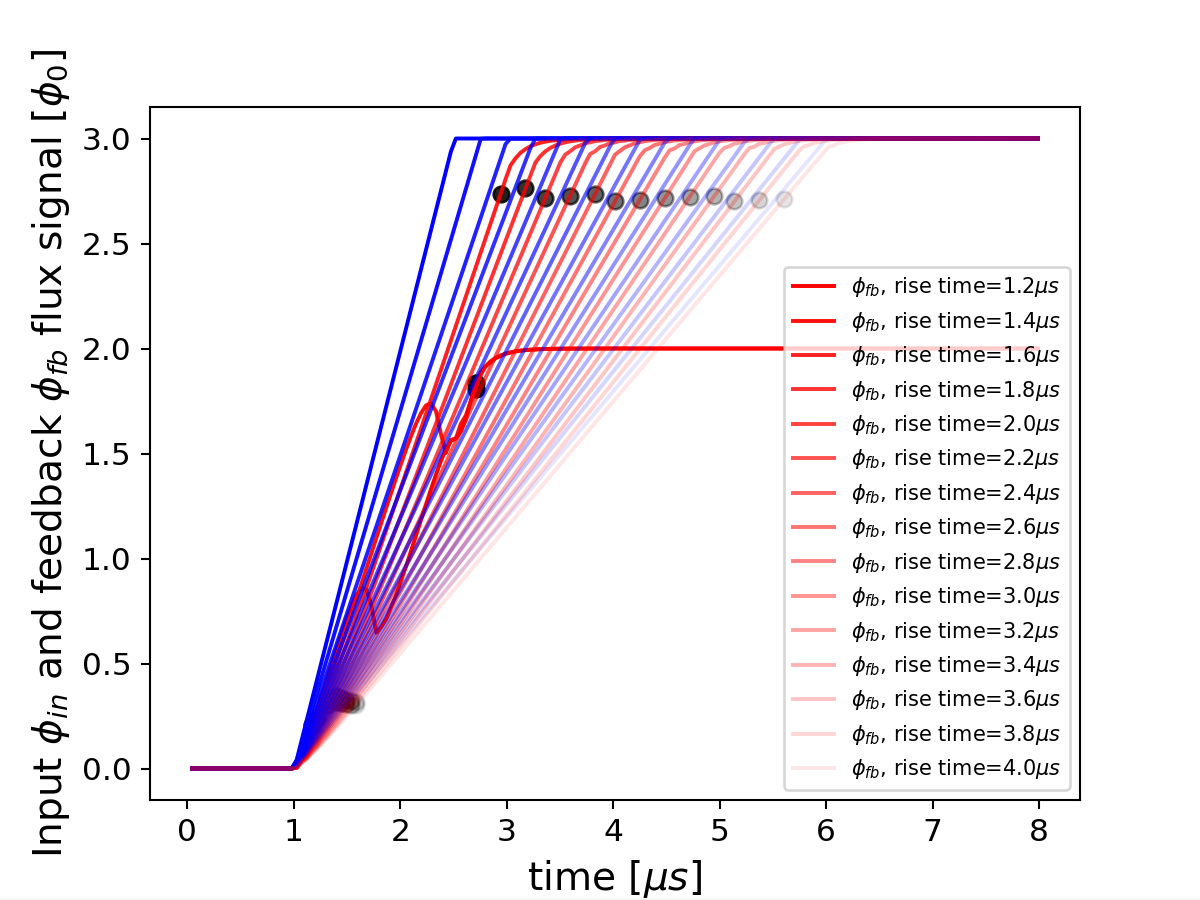}
    \includegraphics[width=.45\textwidth]{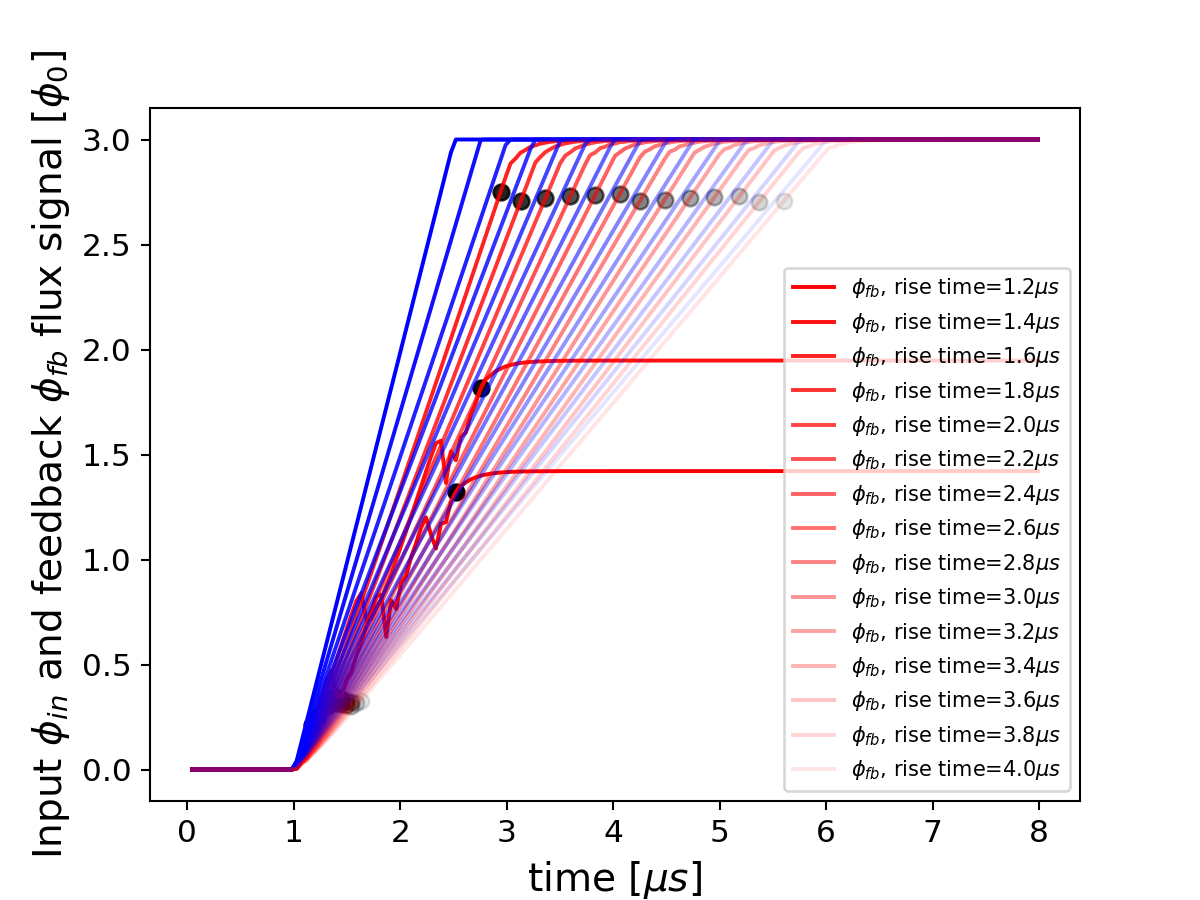}
    \caption{SQUID-FLL system slew rate simulation. The top figure shows the same maximum slew rate of SQUID-FLL system is about $1.6$ $\mathrm{\phi_0/{\mu}s}$ when the SQUID transfer functions are sinusoidal-like model or small bias model with superconducting state. The rise time of $\phi_{fb}$ (red line) is about $1.6$ $\mathrm{{\mu}s}$ and the amplitude of $\phi_{fb}$ is up to 3 $\mathrm{\phi_0}$ without distortion. Rise time is from 10 $\mathrm{\%}$ to 90$\mathrm{\%}$ amplitude of signal. The bottom figure shows the same maximum slew rate of SQUID-FLL system based on asymmetrical model.\label{fig:sim_FLL_lock_philin}}
\end{figure}
Based on Equation~\ref{eq:6}, we calculated $\partial{\phi_{fb}}/\partial{t}{\approx}1.5$ $\mathrm{\phi_0/{\mu}s}$. The simulated result ($t_{set}$=$0$) was approximately $1.5$ $\mathrm{\phi_0/{\mu}s}$, as shown in Figure~\ref{fig:sim_FLL_lock_philin}. 
When the signal rise time exceed $1.6$ $\mathrm{{\mu}s}$, the FLL is unable to accurately track the signal, resulting in distortion.
% While varying distortions appear when the SQUID transfer model is different, all signals are distorted at rise times less than 1.6 ${\mu}s$, which is independent of the shape of the SQUID transfer model.
It can be seen that when the SQUID transfer model is different, the shape of the tracked signal is different when distortion appears. But the value of the slew rate is same. All signals are distorted at rise times less than 1.6 ${\mu}s$, which is independent of the shape of the SQUID transfer model.
For digital FLL, the slew rate decreases when the frame frequency is reduced, as shown in Figure~\ref{fig:sim_FLL_lock_frame_clk}.
In theory, to track a faster signal, one can adjust the integration parameters, feedback resistance and so forth to modify the bandwidth. 
According to equations~\ref{eq:5.1} and~\ref{eq:6}, the ideal linear relationship between slew rate and bandwidth is shown in Figure~\ref{fig:sim_FLL_lock_slewrate_fc}.

\subsection{Noise contribution in digital FLL system}
The TES detector has a high energy resolution mainly due to its small thermal noise at low temperatures.
In the experiment, the noise is mainly comprised of TES Johnson noise and thermal fluctuation noise (TFN)~\cite{TES_in_CPD}.
The thermal noise of SQUID and the voltage noise of preamplifier mainly constitute the system's high-frequency noise.
For digital FLL circuits, the quantization noise of the ADC and DAC devices also affects the signal-to-noise ratio (SNR). The noise spectral density $S(f)$ of ADC or DAC should satisfy Equation~\ref{eq:7}.
\begin{equation}
\label{eq:7}
% \left\{
    % \begin{aligned}
        S(f)=\frac{q}{\sqrt{12}}\frac{1}{\sqrt{BW}} 
    % \end{aligned}
% \right.
\end{equation}
$q$ is the least number bit (LSB) voltage of ADC or DAC. $BW$ is the Nyquist bandwidth of the system.
We can calculate the equivalent current noise density of ADC $S_{I,ADC}(f)$ and DAC $S_{I,DAC}(f)$ refer to the input coil of SQUID according to Equation~\ref{eq:7_2} and~\ref{eq:7_3}.
\begin{equation}
\label{eq:7_2}
% \left\{
    % \begin{aligned}
        S_{I,ADC}(f)=\frac{S_{ADC}(f)}{G_{1}V_{\phi}M_{in}}
    % \end{aligned}
% \right.
\end{equation}
\begin{equation}
\label{eq:7_3}
% \left\{
    % \begin{aligned}
        S_{I,DAC}(f)=\frac{S_{DAC}(f)M_{fb}}{R_{fb}M_{in}}
    % \end{aligned}
% \right.
\end{equation}
Where $S_{ADC}(f)$ and $S_{DAC}(f)$ represents the intrinsic noise of ADC and ADC.
For example, the equivalent current noise of a 16-bit ADC with 2 $\mathrm{V}$ full scale range is 0.8 $\mathrm{pA/\sqrt{Hz}}$ refers to the input coil of SQUID, which is far less than the noise of TES (generally several tens to hundreds $\mathrm{pA/\sqrt{Hz}}$). The equivalent noise of a 16-bit DAC is 0.6 $\mathrm{pA/\sqrt{Hz}}$ when the feedback resistor is 10 $\mathrm{k\Omega}$ as shown in Table~\ref{tab:1}.
Thus, it is better to choose a large feedback resistor and high-resolution ADC and DAC for digital FLL.

\subsection{X-ray pulse signal response}
\begin{figure*}[!htb]
    \centering
    \includegraphics[width=.8\textwidth]{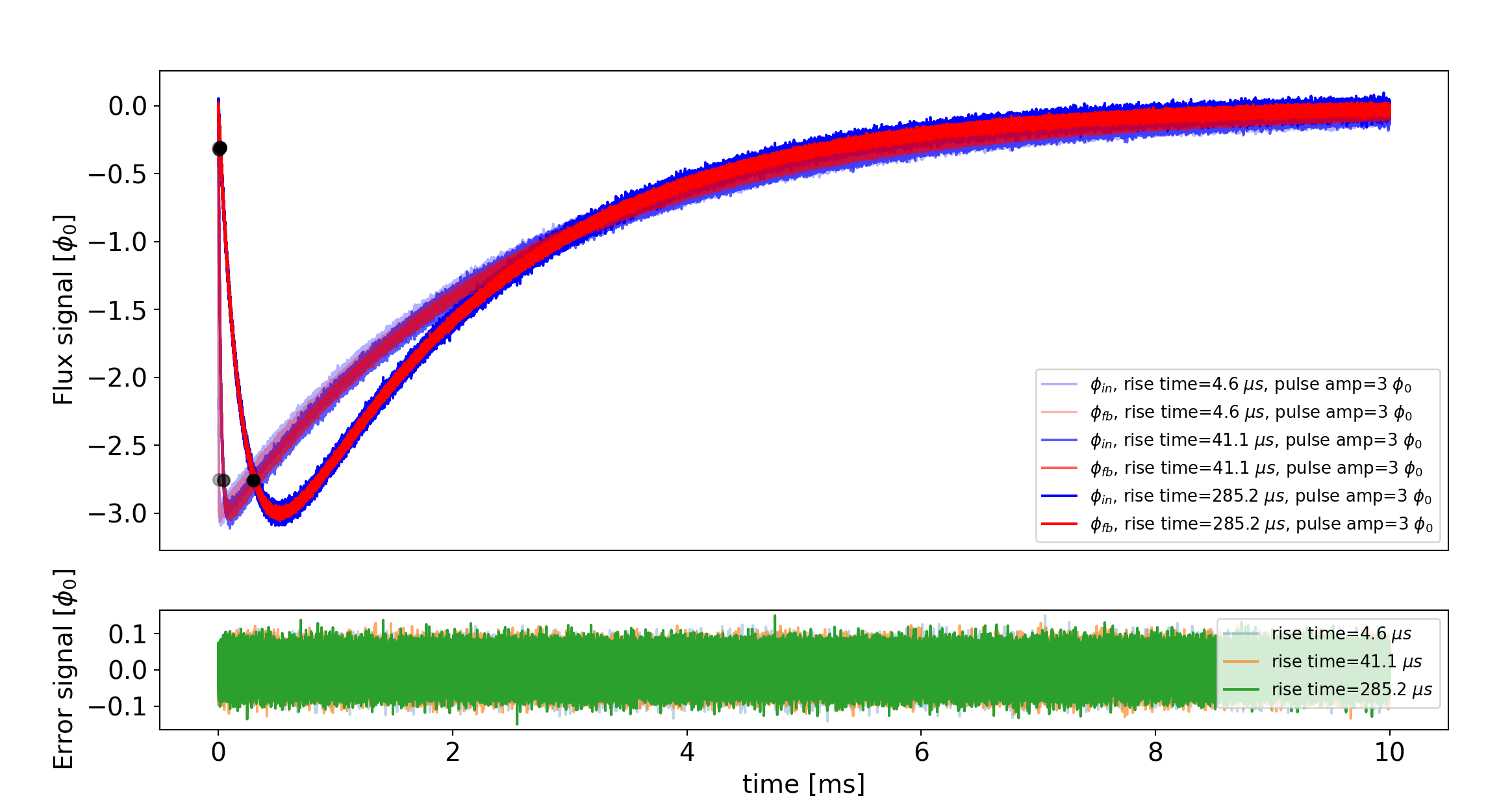}
    \includegraphics[width=.8\textwidth]{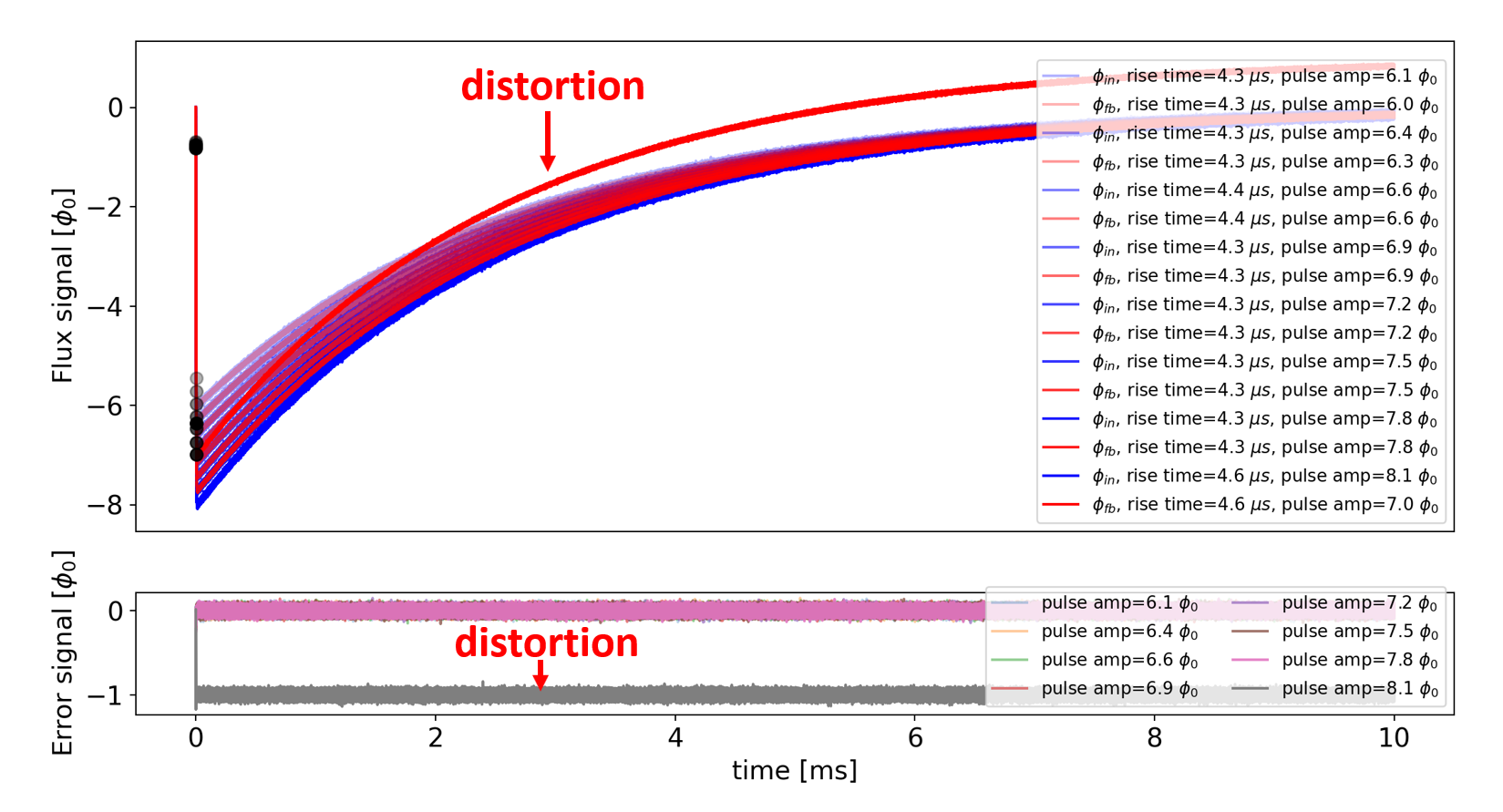}
    \caption{Simulation of X-ray pulses with $22$ $\mathrm{{\mu}A}$ RMS noise. (a) Signal (blue lines) can be tracked well (red lines) when the rise time is not less than $4.3$ $\mathrm{{\mu}s}$ and the amplitude of pulse is $3$ $\mathrm{{\phi_{0}}}$. (b) When the amplitude is larger than $7.8$ $\mathrm{\phi_0}$ and the rise time is about $4.3$ $\mathrm{{\mu}s}$, signal distortion appears.   \label{fig:sim_FLL_lock_pulse}}
\end{figure*}

Considering that the TES used for X-ray detection has a fast response time of the order of a few microseconds to several hundred microseconds, we use a small pulse signal comprising white noise as input to simulate the response process of the X-ray TES\cite{TES_in_CPD}. 
If the Nyquist noise bandwidth is 1 $\mathrm{MHz}$, the noise of the TES system was set to 50 $\mathrm{pA/\sqrt{Hz}}$ at 1 $\mathrm{kHz}$ and an RMS value of 22 $\mathrm{{\mu}A}$ (refer to the SQUID input coil), which is based on our testing results~\cite{Li_2024}.
We use the sinusoidal-like SQUID-FLL model to demonstrate effective tracking of pulses with varying rise time, as illustrated in Figure~\ref{fig:sim_FLL_lock_pulse} (a).
Furthermore, a simulation was performed in which the rise time was set at 4.3 $\mathrm{{\mu}s}$ while the pulse amplitude was increased. The results of this simulation are shown in Figure~\ref{fig:sim_FLL_lock_pulse} (b).
Distortion occurs when the signal amplitude exceeds 7.8 $\mathrm{\phi_0}$. The corresponding slew rate is approximately 1.45 $\mathrm{\phi_0/{\mu}s}$, which is consistent with the simulation results shown on the right side of Figure~\ref{fig:sim_FLL_lock_philin}.

\section{Conclusion}
\label{sec:conclusion}
A comprehensive behavioral simulation of the SQUID-FLL circuit was performed using different SQUID models in conjunction with the principles of digital PID magnetic flux feedback.
We demonstrate that the design of the FLL does not depend on the shape of the SQUID $V$-$\phi$ response by simulating SQUID transfer function models with three different shapes.
A comprehensive list of parameters that must be considered when designing the SQUID-FLL system is presented in Table~\ref{tab:1}.
Common scenarios of locking point adjustments during laboratory tuning were analyzed.
The primary factors influencing the design of the system bandwidth $f_c$ were examined, thereby providing valuable references for the design of SQUIDs and room readout electronics.
% The relationship between $\dot{\phi_{fb}}$ and $f_c$ is consistent with the 
Finally, the upper limit of the slew rate of the system was thoroughly analyzed, and it was found to be approximately 1.5 $\mathrm{\phi_0/{\mu}s}$ under default parameters in our model. 
The tracking capability of the system was validated with simulated X-ray pulse signals with different rise time (4.4 to 281 $\mathrm{{\mu}s}$) and amplitudes (6 to 8 $\mathrm{\phi_0}$).
We hope that these simulation results will serve as a reference for the design of high-performance digital PID flux feedback electronic and multiplexing readout systems.

\bibliographystyle{unsrt}
\bibliography{reference} % bib file

\end{document}